\documentclass[osajnl,twocolumn,showpacs,superscriptaddress,10pt]{revtex4-1} %% use 11pt for Applied Optics
\usepackage{amsmath,amssymb,graphicx}
\usepackage{color,ulem}

\begin{document}

\title{Electromagnetic energy stored in inhomogeneous scattering systems}

\author{\firstname{Tiago}  J. \surname{Arruda}}
\email{tiagoarruda@pg.ffclrp.usp.br}
\affiliation{Instituto de F\'isica de S\~ao Carlos,
Universidade de S\~ao Paulo, 13566-590 S\~ao Carlos, S\~ao Paulo, Brazil}

\author{\firstname{Felipe}  A. \surname{Pinheiro}}
\affiliation{Instituto de F\'{i}sica, Universidade Federal do Rio de Janeiro, 21941-972, Rio de Janeiro-RJ, Brazil}

\author{\firstname{Alexandre} S. \surname{Martinez}}
%\email{asmartinez@ffclrp.usp.br}
\affiliation{Faculdade de Filosofia,~Ci\^encias e Letras de Ribeir\~ao
Preto, Universidade de S\~ao Paulo, 14040-901, Ribeir\~ao
Preto-SP, Brazil}
\affiliation{National Institute of Science and Technology in
Complex Systems, 22290-180 Rio de Janeiro, Rio de Janeiro, Brazil}

%\date{\today}

\begin{abstract}

We analytically study the time-averaged electromagnetic energy stored inside scatterers containing inclusions of arbitrary shapes.
Assuming the low density of inclusions, we derive the expression for the energy-transport velocity through disordered media without relying on the radiative transfer equation.
Moreover, this expression is independent of the shape of scatterers.
In addition, we obtain a relation between the dwell and absorption times associated with inclusions by considering the relationship between internal energy and absorption cross-section.
An approximation for the electromagnetic energy stored inside a disordered medium in terms of the transport mean free path and the packing fraction is also derived.
This expression suggests that the enhanced electromagnetic energy within the host medium is achieved for inclusions exhibiting negative scattering asymmetry parameters.
As a result, disordered media with enhanced backscattering is expected to exhibit large quality factors.

\end{abstract}


\maketitle

%\tableofcontents

\section{Introduction}

In the multiple scattering theory, when the weak disorder approximation and low density of scatterers, quantities calculated in a single-scattering process can be applied to picture the electromagnetic (EM) energy transport through disordered media~\cite{ishimaru}.
Relying on the accuracy of the Lorenz-Mie theory~\cite{bohren}, such approximation provides an analytical approach to estimate multiple scattering parameters without considering full-wave simulations or multiple scattering algorithms, which demand high computational power.
Among the single-scattering properties, the time-averaged EM energy inside the scatterers is a quantity closely related to the photon fluxes in disordered media, finding several applications in photonics and metamaterials~\cite{tiago-pra2016}. 
This quantity can be related to the absorption cross-section~\cite{bott,tiago-sphere,tiago-cylinder} and is associated, e.g., with the dwell-time~\cite{tiggelen}, the quality factor~\cite{pinheiro}, the Verdet constant~\cite{lacoste} and the energy-transport velocity~\cite{albada}, which are relevant measurable quantities for technological applications such as random lasers~\cite{pinheiro} and dispersive metamaterials~\cite{ruppin-energy,tiago-joa,tiago-active,tiago-coated-cylinder}. 

Here, we calculate the stored EM energy inside a single scatterer containing inclusions of arbitrary shapes.
By deriving the relationship between the internal energy and the absorption cross-section, we can define an absorption time associated with a single scattering.
Also, in the incoherent approximation for low density of inclusions, we demonstrate that the energy-transport velocity through a disordered medium can be readily calculated within our approach for scatterers of arbitrary shapes.
Finally, we estimate the EM energy within the host medium to calculate the dwell time related to the system as a function of multiple scattering quantities.
Interestingly enough, we show that a high quality factor for lasing applications can be achieved for scatterers with negative asymmetry parameter, $\langle\cos\theta\rangle<0$.    
This suggests that particles satisfying, e.g., the second Kerker condition ($\langle\cos\theta\rangle=-1/2$), which exhibits near-zero forward scattering response~\cite{nieto}, or exhibiting strong backscattering response ($\langle\cos\theta\rangle=-1$) may lead to large quality factors associated with the disordered medium. 
This multiple scattering regime is referred to as anomalous scattering, since the transport mean free path $\ell_{\rm tr}$ is smaller than the scattering mean free path $\ell_{\rm sca}$~\cite{gomez}.

\section{Time-averaged electromagnetic energy}

\begin{figure}[htbp]
\centerline{\includegraphics[width=\columnwidth]{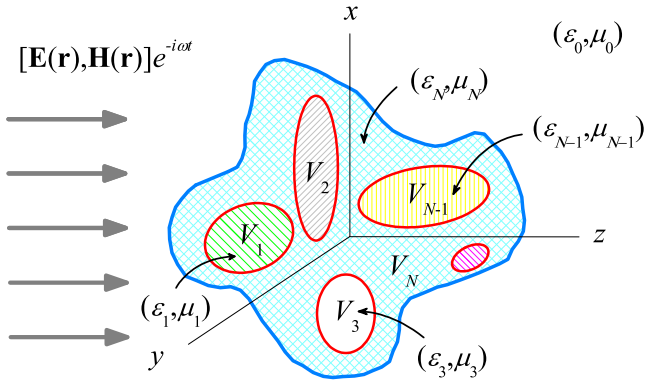}}
\caption{Scatterer with an arbitrary shape and $(N-1)$ arbitrary inclusions irradiated by a plane EM wave.
The surrounding medium is ($\varepsilon_0,\mu_0)$ and each inclusion has volume $V_q$ and optical properties $(\varepsilon_q,\mu_q)$, with $q=\{1,\ldots,N-1\}$, being $V_N$ the volume of the host particle $(\varepsilon_N,\mu_N)$.}\label{fig1}
\end{figure}

Let us consider a composite medium made of a host of volume $V$ and $(N-1)$ inclusions of arbitrary shapes ($N\geq1$ is an integer number), as depicted in Fig.~\ref{fig1}.
Each inclusion has delimited and well-defined volumes $V_1, V_2, \ldots, V_{N-1}$, such that the scatterer volume is $V=\sum_{q=1}^N V_q$, with $V_N$ being the host particle volume ({i.e.}, the scatterer volume minus inclusions).
These $N$ inclusions are linear, homogeneous and isotropic, with electric permittivity $\varepsilon_q=\varepsilon_q'+\imath\varepsilon_q''$ and magnetic permeability $\mu_q=\mu_q'+\imath\mu_q''$ associated with each volume $V_q$.
This scatterer is irradiated by a plane EM wave with electric amplitude $E_0$ and time dependency $e^{-\imath\omega t}$, where $\omega$ is the input frequency and $H_0=\sqrt{\varepsilon_0/\mu_0}E_0$ is the magnetic amplitude.
The surrounding medium is vacuum $(\varepsilon_0,\mu_0)$.
Given the EM field $(\mathbf{E}_{q},\mathbf{H}_{q})$ at position $\mathbf{r}$, confined inside a passive inclusion $(\varepsilon_q,\mu_q)$ with volume $V_q$, the time-averaged EM energy density~\cite{tiago-sphere,tiago-cylinder,tiago-joa,tiago-active,tiago-coated-cylinder} is
\begin{align}
 \langle u_q\rangle_t(\mathbf{r}) = \varepsilon_q^{\rm(eff)}\left|\mathbf{E}_q(\mathbf{r})\right|^2+\mu_q^{\rm(eff)}\left|\mathbf{H}_q(\mathbf{r})\right|^2,
 \label{energy_density}
\end{align}
where $[\varepsilon_q^{\rm(eff)},\mu_q^{\rm(eff)}]$ are energy coefficients associated with the medium $q=\{1,2,\ldots,N\}$ and the time-averaged EM energy in the inclusion is
\begin{align}
    W_q=\frac{1}{4} \; \int_{V_q}{\rm d}^3r {\langle u_q\rangle_t(\mathbf{r})} .
    \label{Wqt}
\end{align}
If the medium $(\varepsilon_q,\mu_q)$ is weakly absorbing ($|\varepsilon_q''|\ll\varepsilon_q'$, $|\mu_q''|\ll\mu_q'$), then $\varepsilon_q^{\rm(eff)}=\partial(\omega\varepsilon_q')/\partial\omega>0$ and $\mu_q^{\rm(eff)}=\partial(\omega\mu_q')/\partial\omega>0$~\cite{landau}.
Also, if the region $V_q$ has the same optical properties as the surrounding medium $(\varepsilon_0,\mu_0)$, it follows that $W_{0q} = (\varepsilon_0|E_0|^2+\mu_0|H_0|^2)V_q/4$, so that~\cite{bott}:
\begin{align}
    W_{0q} =\frac{\varepsilon_0}{2}|E_0|^2V_q .\label{Wq0}
\end{align}

From Eqs.~(\ref{Wqt}) and (\ref{Wq0}), it is clear that the normalized time-averaged energy $W_q/W_{0q}$ is a spatial average ($\langle\cdots \rangle_{V_q} = V_q^{-1}\int_{V_q}{\rm d}^3r \cdots$) over time-averaged field intensities within the inclusion $(\varepsilon_q,\mu_q)$.
Defining the volume-averaged field intensity $\langle |\mathbf{F}_q/F_0|^2\rangle_{V_q} = V_q^{-1}\int_{V_q}{\rm d}^3r |\mathbf{F}_q/F_0|^2$, where $\mathbf{F}_q/F_0$ can be either $\mathbf{E}_q/E_0$ or $\mathbf{H}_q/H_0$, it leads to~\cite{tiago-joa,tiago-coated-cylinder}
\begin{align}
\frac{W_q}{W_{0q}} = \frac{1}{2} \left[\frac{\varepsilon_q^{\rm(eff)}}{\varepsilon_0} \;  \left\langle\left|\frac{\mathbf{E}_q}{E_0}\right|^2\right\rangle_{V_q} + \frac{\mu_q^{\rm(eff)}}{\mu_0} \; \left\langle\left|\frac{\mathbf{H}_q}{H_0}\right|^2\right\rangle_{V_q} \right] ,\label{Wq}
\end{align}
which is the EM energy stored inside the inclusion $(\varepsilon_q,\mu_q)$.
Since the energy $W_q$ is an extensive quantity, it follows that
\begin{align}
    \frac{W}{W_0}= \sum_{q=1}^N \frac{V_q}{V} \; \frac{W_q}{W_{0q}}
     \label{W}
\end{align}
is the total stored EM energy, where $W_0=\sum_{q=1}^N W_{0q}$.

From Eqs. (\ref{Wq}) and (\ref{W}) one can see that the internal energy $W/W_0$ is a weighted arithmetical mean between the electric and magnetic contributions.
Each inclusion $(\varepsilon_q,\mu_q)$ contributes to the sum weighted by its respective volume fraction $V_q/V$.
Note that we are not summing field intensities but time- and volume-averaged field intensities ($W_q$), which are extensive quantities.
This means one may have field interferences inside each region $q$ due to the interaction between inclusions close to each other.
Indeed, owing to boundary conditions provided by the Maxwell's equations, one has that the EM field $(\mathbf{E}_q,\mathbf{H}_q)$ within the inclusion $(\varepsilon_q,\mu_q)$ depends on the resulting field inside the others $(N-2)$ inclusions and the host medium $(\varepsilon_N,\mu_N)$: $\mathbf{E}_q=\mathbf{E}_q(\mathbf{E}_1,\ldots,\mathbf{E}_{q-1},\mathbf{E}_{q+1},\ldots,\mathbf{E}_N)$. 
Hence, determining the EM energy density in each region of the scatterer is usually difficult for general shapes and positions of densely packed inclusions due to multiple scattering interferences in the host medium.
In general, an analytic solution is only possible for scatterers with high degree of symmetry ({e.g.}, spheres and cylinders) and for low density of inclusions. 
Approximations are usually obtained by using, {e.g.}, the Rayleigh-Gans approach~\cite{ishimaru,bohren}.

To analytically proceed let us consider, for the sake of simplicity, two particular geometries with closed solutions: center-symmetric coated spheres and cylinders $(N=2)$.
The absorption efficiency $Q_{\rm abs}$ (which is the absorption cross-section $\sigma_{\rm abs}$ in units of the geometrical one $\sigma_{\rm g}$) can be written in terms of the internal EM field intensities.
For coated spheres and cylinders, both with outer radius $R$: $Q_{\rm abs}=\xi kR\sum_{q=1}^2\langle(\varepsilon_q''/\varepsilon_0) |\mathbf{E}_q/E_0|^2 + (\mu_q''/\mu_0)|\mathbf{H}_q/H_0|^2\rangle_{V_q} (V_q/V)$, where $kR$ is the size parameter, with $k=\omega/c_0$, and $\xi=4/3$ for spheres and $\xi=\pi/2$ for cylinders~\cite{tiago-joa,tiago-coated-cylinder}.
One can show that $\sigma_{\rm g}\xi kR= k V$, where $V$ is either the volume of a sphere or a finite cylinder with radius $b$.
Using this result, one can eliminate the dependence on $kR$ and $\sigma_{\rm g}$, which allows us to write $\sigma_{\rm abs}$ for a coated particle with an arbitrary shape.
Therefore, for an arbitrary particle with $(N-1)$ inclusions, we find
    \begin{align}
        \sigma_{\rm abs} =  \sum_{q=1}^NkV_q\left[\frac{\varepsilon_q''}{\varepsilon_0} \left\langle\left|\frac{\mathbf{E}_q}{E_0}\right|^2\right\rangle_{V_q} + \frac{\mu_q''}{\mu_0} \left\langle\left|\frac{\mathbf{H}_q}{H_0}\right|^2\right\rangle_{V_q}\right] , \label{sigma-abs}
    \end{align}
which generalizes the well-known relation between power loss and absorption for a collection of scatterers~\cite{ishimaru}.

\subsection{Dwell and absorption times}

To obtain a relation between $W$ and $\sigma_{\rm abs}$, let us consider a weak absorbing and non-dispersive medium, so that $[\varepsilon_q^{\rm(eff)}=\varepsilon_q',\mu_q^{\rm(eff)}=\mu_q']$ into Eq.~(\ref{W}).
By defining the complex quantity $\widetilde{W}/W_0$ by replacing $(\varepsilon_q^{\rm(eff)},\mu_q^{\rm(eff)})$ with  $(\varepsilon_q=\varepsilon_q'+\imath\varepsilon_q'',\mu_q=\mu_q'+\imath\mu_q'')$ in Eq.~(\ref{W}), one has
\begin{align}
\frac{\widetilde{W}}{W_0}=\frac{W}{W_0}+\imath\frac{\sigma_{\rm abs}}{2kV} ,\label{W-complex}
\end{align}
where the real part is the stored EM energy and the imaginary part depends on the absorption cross section.
 The denominator $2kV$ in Eq.~(\ref{W-complex}) can be interpreted as a general geometrical cross section for an arbitrary shape scatterer.
In particular, for spherical and cylindrical scatterers, it takes into account both the geometrical cross-section and the size parameter.
In a dispersive medium $q$, we recall that the relation between $\varepsilon_q'$ and $\varepsilon_q''$ is given by the Kramers-Kronig relations~\cite{bohren}.

The dwell-time $\tau_{\rm d}$ is defined as the time that the EM waves are delayed due to resonances inside a scatterer~\cite{tiggelen}:
\begin{align}
\tau_{\rm d} = \frac{V}{\sigma_{\rm sca}c_0}\frac{W}{W_0} .\label{tau-d}
\end{align}
Defining the complex quantity $\widetilde{\tau}_{\rm d} = V(\widetilde{W}/W_0)/(\sigma_{\rm sca}c_0)$ in the frequency domain, we obtain from Eq.~(\ref{W-complex}):
\begin{align}
\widetilde{\tau}_{\rm d}(\omega)=\tau_{\rm d}(\omega)+\imath\tau_{\rm abs}(\omega),
\end{align}
where we define the absorption or relaxation time $\tau_{\rm abs} \equiv {\sigma_{\rm abs}}/({2\omega\sigma_{\rm sca}})$: $\exp(\imath \omega \widetilde{\tau}_{\rm d})=\exp({\imath\omega\tau_{\rm d}-\omega\tau_{\rm abs}})$.
Since the albedo is ${a_0} = \sigma_{\rm sca}/\sigma_{\rm ext}$, we have
\begin{align}
\tau_{\rm abs}(\omega)=\frac{1-{a_0}(\omega)}{2\omega{a_0(\omega)}} .\label{tau-abs}
\end{align}

For weak absorbing scatterers $(a_0\approx1)$, one can write $W/W_0\approx m'\sigma_{\rm abs}/(2km''V)$, where $m=m'+\imath m''$ is the scatterer effective refractive index~\cite{tiago-cylinder}.
Therefore, from Eq.~(\ref{tau-d}),
\begin{align}
    \tau_{\rm d}(\omega)\approx \lim_{m''\to 0}\frac{1-a_0(\omega)}{2\omega a_0(\omega) }\frac{m'}{m''}\approx\lim_{m''\to0}\frac{1-a_0(\omega)}{2\omega }\frac{m'}{m''} ,\label{tau-d-abs}
\end{align}
which is a well-known result that can be proved rigorously for scalar waves~\cite{tiggelen}.
Thus, for weak absorption, using Eq.~(\ref{tau-abs}), one has
\begin{align}
\tau_{\rm d} \approx \lim_{m''\to0}\left(\frac{m'}{m''}\right)\tau_{\rm abs}=\lim_{v_{\rm p}\ll2\omega\ell_{\rm ext}} \left(\frac{2\omega\ell_{\rm ext}}{v_{\rm p}}\right)\tau_{\rm abs}\ , 
\end{align}
where we have considered $m={c_0}/{v_{\rm p}} +\imath {c_0}/({2\omega\ell_{\rm ext}})$,
with $\ell_{\rm ext}$ being the extinction mean-free path and $v_{\rm p}$ being the phase velocity~\cite{tiggelen}.

In short, the relationship between $\sigma_{\rm abs}$ and $W/W_0$ has allowed us to link the dwell and absorption times associated with an arbitrary scatterer containing inclusions.
Here $\tau_{\rm abs}$ is the absorption time of a disordered system.
As we have already pointed out, analytical calculations for arbitrarily packed inclusions are very cumbersome.
To make our discussions as simple as possible, from now on we consider the incoherent approximation for low density of inclusions, so that each inclusion can be seen as a single independent scatterer. 

\subsection{Energy transport velocity}

\begin{figure}[htbp]
\centerline{\includegraphics[width=\columnwidth]{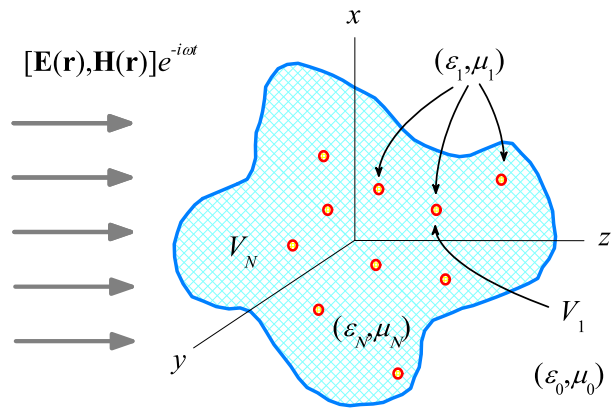}}
\caption{A scatterer with arbitrary shape consisting of $(N-1)$ identical arbitrary inclusions $(\varepsilon_1,\mu_1)$ embedded in a host medium $(\varepsilon_N,\mu_N)$ irradiated by plane EM waves.
The host medium has volume $V_N$ and each inclusion has volume $V_1$, so that $(N-1)V_1\ll V_N$.}\label{fig2}
\end{figure}

The energy-transport velocity $v_{\rm E}$ is defined as the ratio between the time-averaged Poynting vector and the energy density: $v_{\rm E} =|\langle \mathbf{S}\rangle_t|/\langle u\rangle_t$.
Let us consider a particular case with $(N-1)$ identical inclusions of volume $V_1$ and the same optical properties $(\varepsilon_1,\mu_1)$, as depicted in Fig.~\ref{fig2}.
Also, the host particle volume is much larger than the inclusions [$V_N\gg (N-1) V_1$] and each mean inclusion separation is much greater than their radius.
Thus, we can consider that the volume-averaged intensity inside $V$ is approximately the volume-averaged intensity inside $V_N$.
This is valid even in the resonant scattering $\langle|\mathbf{E}_q|^2\rangle_{V_1}\gg |E_0|^2$ provided that the filling fraction of inclusions is very low $[(N-1)V_1/V\ll1]$. 
Considering that $(\varepsilon_N,\mu_N)$ has the same optical properties as the surrounding medium $(\varepsilon_0,\mu_0)$, one has $|\langle \mathbf{S}\rangle_t|\approx E_0H_0/2=c_0\varepsilon_0|E_0|^2 /2=c_0W_0/V$.
From Eq.~(\ref{W}), note that, for identical particles $(\varepsilon_1,\mu_1)$, one can write
\begin{align}
    \frac{W_{\rm bulk}}{W_0}=(N-1)\frac{V_1}{V}\frac{\overline{W}_1}{W_{01}} + \frac{V_N}{V} ,\label{W-app1}
\end{align}
where $\overline{W}_1=\sum_{q=1}^{N-1}W_q/(N-1)$ is the arithmetic mean energy inside the inclusions.
Approximating  $\langle u\rangle_t\approx W/V$ and calling the packing fraction of scatterers $f\equiv(N-1)V_1/V$, then $V_N/V=1-f$, it leads to
\begin{align}
    v_{\rm E}=\frac{|\langle\mathbf{S}\rangle_t|}{\langle u\rangle_t}\approx \frac{c_0}{1+f(\overline{W}_1/W_{01}-1)} ,\label{ve}
\end{align}
which is the same result found in Ref.~\cite{albada} using the Bethe-Salpeter equation under weak disorder approximation.
Indeed, if all inclusions $(\varepsilon_1,\mu_1)$ have the same orientation with regard to the incident EM field ({i.e.}, they can be obtained by translation of any other inclusion), one has $\overline{W}_1=W_1=W_q$ for $q=\{1,\ldots,N-1\}$, which is in agreement with Ref.~\cite{albada}.
In other words, by simple assumptions, we have determined the energy-transport velocity in a disordered medium with low density of identical scatterers (inclusions) with arbitrary shapes.
Our heuristic approach does not take directly into account the disorder average as it is usually the case in rigorous multiple scattering calculations~\cite{ishimaru}.
It should also be emphasized that Eq.~(\ref{ve}) does not impose any restriction regarding the incident wavelength and inclusions size~\cite{albada}.
This implies that each inclusion can be described by the generalized Lorenz-Mie theory~\cite{albada,ruppin}.

\subsection{Quality factor}

To calculate the energy-transport velocity through a disordered medium with low density of scatterers, we have neglected the scattered EM fields within $V_N$, the host medium.
This assumption is supported by performing the disorder average, which eliminates the overall single-scattered fields~\cite{albada}.
As a result, for a very low density of scatterers with $f\ll 1$, one has $W=f \sum_{q=1}^{N-1}{W}_q + (1-f)W_N\approx W_N\approx W_0$, provided that $f \sum_{q=1}^{N-1}{W}_q\ll W_N$. 
The last condition, however, has been shown to be not mandatory and is usually not satisfied in a resonant scattering process~\cite{ruppin}. 

To obtain a better approximation for $W$ in a disordered system, we must consider the total electric field in the host medium $(\varepsilon_N=\varepsilon_0,\mu_N=\mu_0)$, which contains both the incident and scattered electric fields.
To simplify our discussion, let us consider that the inclusions are spherical particles with mean radius $R$ and volume $V_1$.
The total electric field within the host medium is
\begin{align}
\mathbf{E}_N(\mathbf{r}) = E_0 e^{\imath kz}\hat{\mathbf{p}}+E_0\sum_{q=1}^{N-1}\frac{\mathbf{f}(\hat{\mathbf{k}}_{q},\hat{\mathbf{k}})}{\imath k r_q}e^{\imath kr_q} ,\label{E_N}
\end{align}
where $\hat{\mathbf{p}}$ is a unit (polarization) vector, $r_q=|\mathbf{r}-\mathbf{r}_q|$, with $\mathbf{r_q}$ being the position of the inclusions, $k=|\mathbf{k}|$ is the incident wave number, and $\mathbf{f}(\hat{\mathbf{k}}_q,\hat{\mathbf{k}})$ is the scattering amplitude related to inclusion $(\varepsilon_q,\mu_q)$.
The energy flux per unit of area is given by the time-averaged Poynting vector: $\langle \mathbf{S}_{N}\rangle_t={\rm Re}(\mathbf{E}_{N}\times\mathbf{H}_{N}^*)/2=c_0\varepsilon_0|\mathbf{E}_{N}|^2\hat{\mathbf{k}}_{N}/2$.

Neglecting interference among multiple scattered fields, one has
\begin{align}
\langle \mathbf{S}_N\rangle_t &= \langle\mathbf{S}_{\rm inc} \rangle_t + \sum_{q=1}^{N-1}\langle\mathbf{S}_{\rm sca}^{(q)} \rangle_t + \mathop{\sum_{q=1}^{N-1}\sum_{{q'=1}}^{N-1}}_{(q\not=q')}\langle\mathbf{S}_{\rm sca}^{(q,q')} \rangle_{t} +\ldots\nonumber\\
&\approx \langle\mathbf{S}_{\rm inc} \rangle_t + \sum_{q=1}^{N-1}\langle\mathbf{S}_{\rm sca}^{(q)} \rangle_t \nonumber\\
&=c_0\varepsilon_0\frac{|E_0|^2}{2}\left[\hat{\mathbf{k}} + \sum_{q=1}^{N-1}\left|\frac{\mathbf{f}(\hat{\mathbf{k}}_q,\hat{\mathbf{k}})}{kr_q}\right|^2 \hat{\mathbf{k}}_q\right] .\label{S-N}
\end{align}
To determine the EM energy in the host medium, we have to eliminate the vector character of Eq.~(\ref{S-N}).
%Considering $|\langle\mathbf{S}_N\rangle_t|^2$, we obtain
%\begin{align}
%\left|\frac{\mathbf{E}_N}{E_0}\right|^4 &\approx 1 + 2\sum_{q=1}^{N-1}\left|\frac{\mathbf{f}(\hat{\mathbf{k}}_q,\hat{\mathbf{k}})}{kr_q}\right|^2 \hat{\mathbf{k}}_q\cdot\hat{\mathbf{k}}\nonumber\\
%&+\sum_{q=1}^{N-1}\sum_{q'=1}^{N-1}\left|\frac{\mathbf{f}(\hat{\mathbf{k}}_q,\hat{\mathbf{k}})}{kr_q}\right|^2\left|\frac{\mathbf{f}(\hat{\mathbf{k}}_{q'},\hat{\mathbf{k}})}{kr_{q'}}\right|^2 \hat{\mathbf{k}}_{q}\cdot\hat{\mathbf{k}_{q'}}\ .\label{Sn}
%\end{align}
However, instead of considering $|\langle\mathbf{S}_N\rangle_t|$ directly, which leads to a cumbersome expression to be integrated, we simply rewrite Eq.~(\ref{S-N}) and multiply it by $\hat{\mathbf{k}}$, i.e., we project $\langle\mathbf{S}_N\rangle_t$ onto the forward direction:
\begin{align}
\left|\frac{\mathbf{E}_N}{E_0}\right|^2 \hat{\mathbf{k}}_N\cdot\hat{\mathbf{k}}\approx 1 + \sum_{q=1}^{N-1}\left|\frac{\mathbf{f}(\hat{\mathbf{k}}_q,\hat{\mathbf{k}})}{kr_q}\right|^2 \hat{\mathbf{k}}_q\cdot\hat{\mathbf{k}} .\label{Sn}
\end{align}
This procedure allows us to keep the dependence of the intensity $|\mathbf{E}_N|^2$ on the scattering directions. 
In fact, now we can define $\cos\theta_N\equiv \hat{\mathbf{k}}_N\cdot\hat{\mathbf{k}}$ and $\cos\theta_q\equiv \hat{\mathbf{k}}_q\cdot\hat{\mathbf{k}}$, where $\theta_q$ is the scattering angle related to inclusion $(\varepsilon_q,\mu_q)$.
Calculating the volume average of Eq.~(\ref{Sn}) over $V_N$ and recalling the well-known definition of the asymmetry parameter $\langle\cos\theta_q\rangle$~\cite{bohren}, we obtain
\begin{align}
\left\langle \left|\frac{\mathbf{E}_N}{E_0}\right|^2 \cos\theta_N\right\rangle_{V_N} &\approx 1 + \frac{R}{V_N}\sum_{q=1}^{N-1}\int_{4\pi} {\rm d}\Omega_q \frac{|\mathbf{f}(\hat{\mathbf{k}}_q,\hat{\mathbf{k}})|^2}{k^2}\cos\theta_q\nonumber\\
&= 1 + \frac{R}{V_N}\sum_{q=1}^{N-1} \sigma_{\rm sca}^{(q)}\langle \cos\theta_q\rangle\nonumber\\
&= 1 + \frac{R\rho \overline{\sigma}_{\rm sca}\langle\cos\theta\rangle}{1-f} ,\label{intensity}
\end{align}
where $\rho = (N-1)/V$ is the density of inclusions, $1-f=V_N/V$, with$f$ being the packing fraction, and $R$ is the mean radius of the spherical scatterers.
Once again, we do not take directly into account the disorder average, since our aim is to provide simple heuristic expressions using only the low density of scatterers and incoherent approximations.
As demonstrated in the last section, these approximations with time- and volume-averages over the field intensities can capture the physical picture. 
It is worth mentioning that an identical expression can be obtained considering parallel cylindrical scatterers of radius $R$ normally irradiated by plane waves instead of spherical scatterers.
One can easily prove that by rewriting Eq.~(\ref{E_N}) for cylindrical scattered waves and following the same procedure above.
Also, since we have different orientations of inclusions in relation to $\hat{\mathbf{k}}$, even for independent and identical scatterers, we have defined the weighted average quantities~\cite{Grenfell}: $\overline{\sigma}_{\rm sca}\langle\cos\theta\rangle\equiv\sum_{q=1}^{N-1}\sigma_{\rm sca}^{(q)}\langle\cos\theta_q\rangle/(N-1)$.
Of course, for spherical or cylindrical inclusions, it follows: $\overline{\sigma}_{\rm sca}=\sigma_{\rm sca}^{(q)}$ and $\langle\cos\theta\rangle=\langle\cos\theta_{q}\rangle$, for $q=\{1,\ldots,N-1\}$.

Equation~(\ref{intensity}) provides the intensity enhancement in the forward direction in the host medium: $|\mathbf{E}_N/E_0|^2\cos\theta_N$. 
The angle $\theta_{N}$ takes into account not only the scattering angles $\theta_q$ related to inclusions, but also the forward incident direction [see Eq.~(\ref{E_N})]. 
Since the ensemble asymmetry parameter is $\langle\cos\theta\rangle$, we consider the following approximation: $\langle\cos\theta_N\rangle\approx 1 + f\langle\cos\theta\rangle/(1-f)$.
This is the simplest approximation for $\langle\cos\theta_N\rangle$ and can be understood as a weighted mean value between the cosine of forward and scattering directions.
It is obtained from Eq.~(\ref{intensity}) by taking $|\mathbf{E}_N/E_0|^2\approx1$ and $R\rho\overline{\sigma}_{\rm sca}\approx f$.
Considering now
$\langle|\mathbf{E}_N/E_0|^2\cos\theta_N\rangle_{V_N}\approx \langle|\mathbf{E}_N/E_0|^2\rangle_{V_N}\langle\cos\theta_N\rangle$, we finally obtain an expression for the intensity enhancement in the host medium.
Also, for plane waves, one has $\langle|\mathbf{E}_N/E_0|^2\rangle_{V_N}=\langle|\mathbf{H}_N/H_0|^2\rangle_{V_N}$. 
Therefore, from Eqs.~(\ref{Wq}) and (\ref{intensity}), and using the aforementioned approximation for $\cos\theta_{N}$, we obtain
\begin{align}
\frac{W_N}{W_{0N}} \approx  \left|\frac{1-f +  R\rho \overline{\sigma}_{\rm sca}\langle\cos\theta\rangle}{1-f(1-\langle\cos\theta\rangle)}\right|.\label{W_N}
\end{align}
Using Eq.~(\ref{W}) and (\ref{intensity}), we finally have
\begin{align}
\frac{W}{W_0}&\approx f\frac{\overline{W}_1}{W_{01}} + \left|\frac{(1-f)\ell_{\rm tr} +  (\ell_{\rm tr}-\ell_{\rm sca})R/\ell_{\rm sca}}{(\ell_{\rm tr} - f\ell_{\rm sca})/(1-f)}\right|,\label{W-main}
\end{align}
where $\overline{W}_1$ is the mean EM energy inside the $(N-1)$ inclusions, $\ell_{\rm sca}=1/\rho\overline{\sigma}_{\rm sca}$ and $\ell_{\rm tr}=\ell_{\rm sca}/(1-\langle\cos\theta\rangle)$ are the scattering and transport mean free paths, respectively.
Equation~(\ref{W-main}) provides an approximation for the EM energy stored inside a system of volume $L^3$ with low density of scatterers as a function of measurable, multiple scattering quantities: $f$, $\ell_{\rm tr}$ and $\ell_{\rm sca}$.
Note that one can retrieve Eq.~(\ref{W-app1}) by considering $\langle\cos\theta\rangle=0$, i.e., isotropic scattering, or $f\ll1$ and $R/\ell_{\rm sca}\ll1$. 

Due to average on disorder and low density of scatterers ($f\ll1$), one can in general assume $W_N\approx W_{0N}$ in random media even at a scattering resonance $(\overline{W}_1\gg W_{01})$.
This implies that the approximation for the energy transport velocity calculated in Eq.~(\ref{ve}) still holds~\cite{albada}. 
However, notwithstanding the incoherent approximation, Eq.~(\ref{W_N}) allows us to study variations on $W_N$ as a function of $f$, $\ell_{\rm tr}$ and $\ell_{\rm sca}$, which become crucial as one increases the density of particles in the system.
Indeed, there are features  in $W_N$ that do  show up in the stored EM energy associated with the whole system, $W$.
This quantity not only provides a correction for $W$ but also unveils properties of the scattered light in the host medium using simple arguments. 

In addition, as the bulk scattering cross-section for $(N-1)$ independent and identical scatterers is $\sigma_{\rm sca}^{\rm (bulk)} =\sum_{q=1}^{N-1}\sigma_{\rm sca}^{(q)}=(N-1)\overline{\sigma}_{\rm sca}$, we can now calculate the bulk dwell-time: $\tau_{\rm d}^{\rm (bulk)}=V[W/W_0]/[\sigma_{\rm sca}^{\rm (bulk)} c_0]=[W/W_0](\ell_{\rm sca}/c_0)$.
Using Eq.~(\ref{W-main}), we obtain the dwell-time for the disordered system in the incoherent approximation:
\begin{align}
\tau_{\rm d}^{\rm(bulk)} = \tau_{\rm d}^{\rm(inclusion)} + \tau_{\rm d}^{\rm(host)} ,\label{tau-d-bulk}
\end{align}
where $\tau_{\rm d}^{\rm(inclusion)}=V_1[\overline{W}_1/W_{01}]/[\overline{\sigma}_{\rm sca} c_0]$ is the mean dwell-time related to one inclusion and
\begin{align}
\tau_{\rm d}^{\rm(host)}\approx \frac{\ell_{\rm sca}}{c_0}\left|\frac{(1-f)\ell_{\rm tr} +  (\ell_{\rm tr}-\ell_{\rm sca})R/\ell_{\rm sca}}{(\ell_{\rm tr} - f\ell_{\rm sca})/(1-f)}\right|.\label{tauu}
\end{align}
For a very dilute disorder medium, with $f\ll1$ and $R/\ell_{\rm sca}\ll1$, one readily obtains $\tau_{\rm d}^{\rm(host)}\approx \ell_{\rm sca}/c_0$, which is an expected result.

In disordered systems, the quality factor $\mathcal{Q}$ measures the ability of a system to store energy for lasing applications, and can be defined as~\cite{pinheiro} 
\begin{align}
\mathcal{Q}\equiv \omega\tau_{\rm d}^{\rm (bulk)}.\label{Q-factor}
\end{align} 
From Eq.~(\ref{tauu}), considering $\overline{W}_1/W_{01}\gg1$ (resonance), a large quality factor can be achieved from the host medium contribution when $f\approx\ell_{\rm tr}/\ell_{\rm sca}$ and $R\approx \ell_{\rm tr}$.
Since $0<f<1$, this implies $\langle\cos\theta\rangle<0$, i.e., preferential backscattering or anomalous scattering regime, with $\ell_{\rm tr}<\ell_{\rm sca}$.

To explain the connection between preferential backscattering in a low density disordered medium and  large quality factors it is useful to consider the dipole approximation.
For dipolar scatterers, preferential backscattering $\langle\cos\theta\rangle<0$ is a result of the interference between electric and magnetic dipole resonances~\cite{nieto,liu}. 
When the electric and magnetic polarizabilities of the scatterers have comparable strength, the so-called second Kerker condition can be fulfilled, leading to almost-zero-forward scattering (with $\langle\cos\theta\rangle=-1/2$)~\cite{gomez}. 
As a result, dilute suspension of such particles near $\langle\cos\theta\rangle=-1/2$ will minimize $\ell_{\rm tr}$ below $\ell_{\rm sca}$. 
Taking Eqs.~(\ref{W_N})--(\ref{Q-factor}) into account, it can be seen that $\langle\cos\theta\rangle<0$ implies in large quality factors.  

Interestingly enough, for non-absorbing particles in the dipolar Mie scattering regime, the ratio $\ell_{\rm tr}/\ell_{\rm sca}=1/(1-\langle\cos\theta\rangle)$ is exactly the ratio between the sum of pure dipole forces and the total radiation force~\cite{gomez}.
From Eq.~(\ref{Q-factor}), a disordered system with $f\approx 0.5$ and $\langle\cos\theta\rangle\approx-1$ is expected to exhibit a large quality factor.
This is also the condition for the maximum magneto-electric radiation force contribution within a disorder medium~\cite{gomez}. 
Recalling that the total radiation force is proportional to EM energy density inside the inclusions~\cite{bohren}, the asymmetry parameter $\langle\cos\theta\rangle<0$ also implies an energy enhancement within the scatterers, then contributing to the quality factor. 

\begin{figure}[htbp]
\includegraphics[width=\columnwidth]{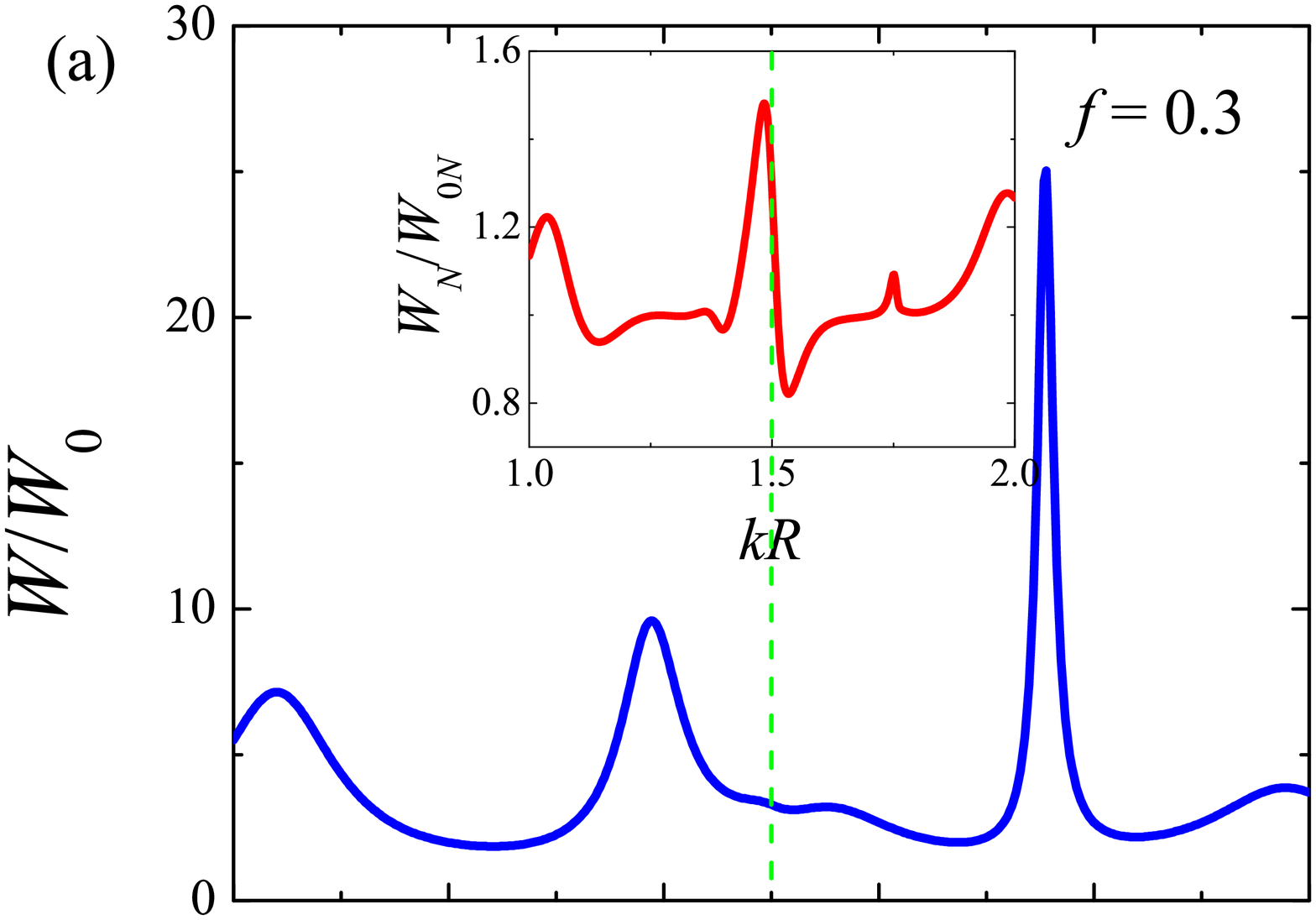}\vspace{-1.8cm}
\includegraphics[width=\columnwidth]{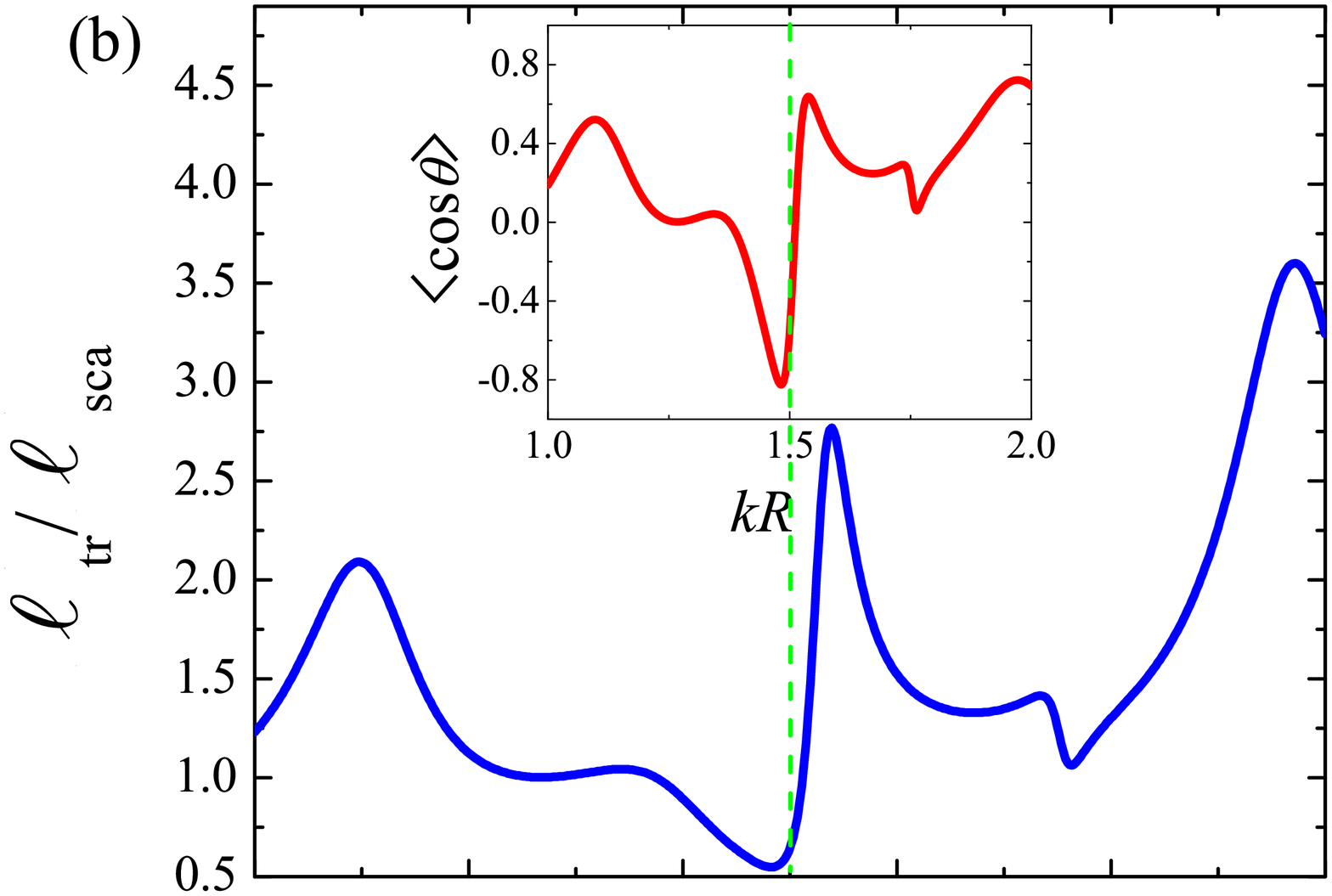}\vspace{-1.8cm}
\includegraphics[width=\columnwidth]{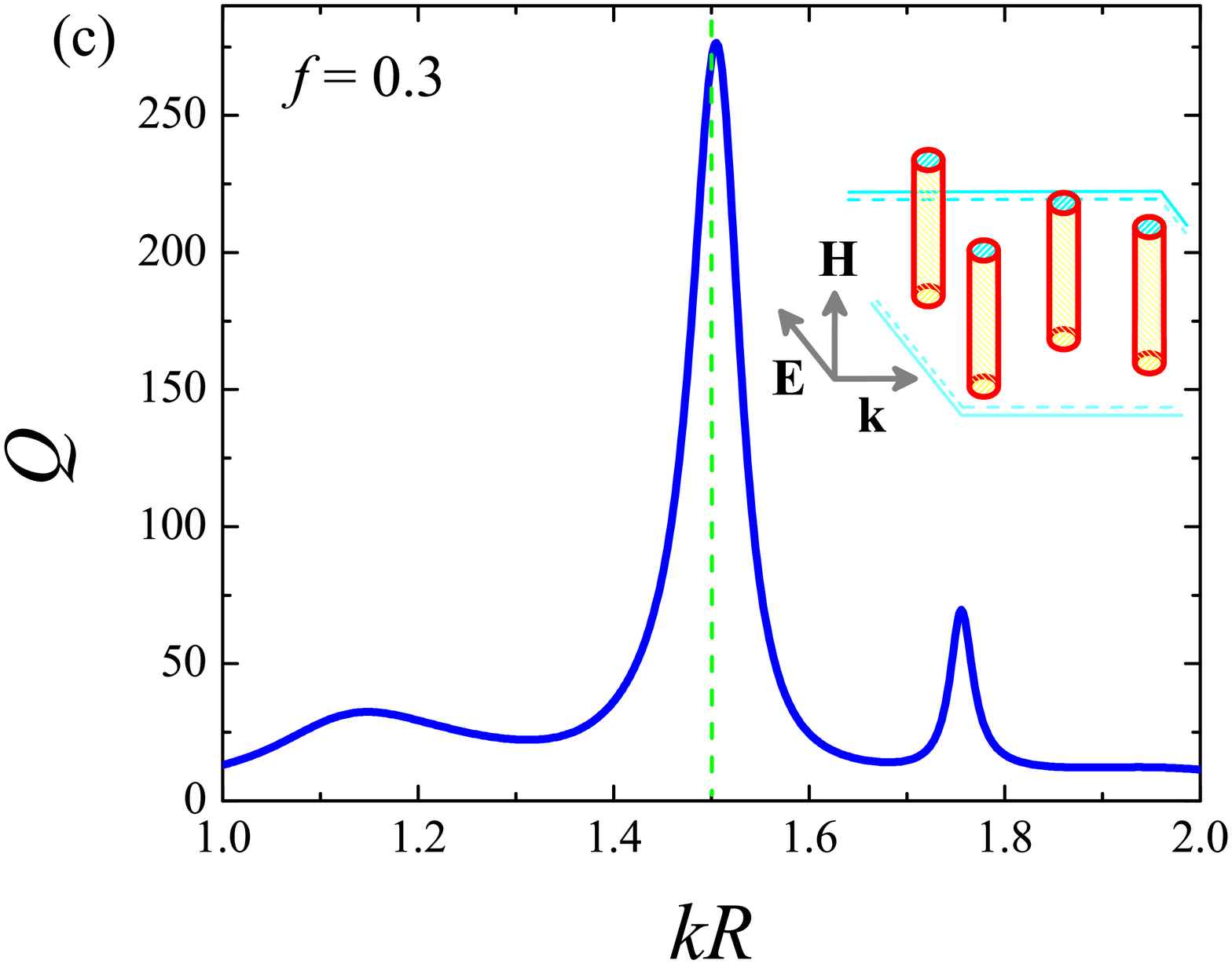}
\caption{A two-dimensional disordered medium consisting of identical parallel cylinders normally irradiated by plane p-waves with $\lambda=395$~nm.
The dielectric cylinders have refractive index $m_1=3.5$ and radius $R$, with a filling fraction $f=0.3$.
(a) The plot shows the total energy enhancement factor $W/W_0$ within the disordered medium as a function of the size parameter $kR$.
The energy stored in the host medium is shown in the inset.
(b) The ratio between transport and scattering mean free paths, $\ell_{\rm tr}/\ell_{\rm sca}$, plotted as a function of $kR$.
The inset shows the scattering asymmetry parameter $\langle\cos\theta\rangle$.
(c) The quality factor $\mathcal{Q}$ of the system plotted as a function of $kR$.
The inset illustrates the geometry of the system. 
The maximum $\mathcal{Q}$ occurs for $kR\approx 1.5$, which approximately corresponds to the maximum energy $W_N/W_{0N}$ in the host medium, with $\ell_{\rm tr}/\ell_{\rm sca}\approx 0.6$ and $\langle\cos\theta\rangle\approx-0.5$.}\label{fig3}
\end{figure}

In Fig.~\ref{fig3}, we study a disordered system in which our approximations are valid.
We consider a disordered array of identical, infinitely long cylinders normally irradiated by p-waves, i.e., the the magnetic field is parallel to the cylinder axis.
The filling fraction is $f=0.3$ and the refractive index of each cylinder is $m_1=\sqrt{\varepsilon_1/\varepsilon_0}=3.5$, with radius $R$.
The quantities $W_1/W_{01}$, $\langle\cos\theta\rangle$, and $\ell_{\rm sca}$ are calculated using the rigorous Lorenz-Mie theory~\cite{tiago-pra2016,tiago-coated-cylinder}. 

The plots in Figs.~\ref{fig3}(a) and \ref{fig3}(b) show that the maximum energy stored in the host medium occurs for $\langle\cos\theta\rangle\approx-0.5$ ($kR\approx 1.5$), i.e., $\ell_{\rm tr}/\ell_{\rm sca}\approx0.6$.
At $kR\approx1.5$, we have satisfied the almost-zero-forward scattering condition due to the destructive interference between electric and magnetic dipole responses. 
Although $kR\approx1.5$ does not correspond to the maximum energy enhancement within the system $(kR\approx1.8$), it corresponds to the maximum quality factor $\mathcal{Q}$, as showed in Fig.~\ref{fig3}(c).
The maximum quality factor for dielectric scatterers can be therefore achieved in the anomalous multiple scatting regime, in which $\ell_{\rm tr}<\ell_{\rm sca}$. 
This result can be readily obtained from the EM energy stored in the host medium, as showed in Fig.~\ref{fig3}(a).

\section{Conclusion}

We have shown some relations between the time-averaged EM energy inside scatterers and multiple scattering quantities in the low density approximation.
Using simple assumptions, we have demonstrated that the energy-transport velocity through a disordered medium can be obtained from the EM energy stored in a single particle containing identical inclusions.
By the relationship between the internal energy and the absorption cross section, we have derived a relation between dwell and absorption times in a disordered medium.
In addition, under the incoherent approximation for the internal EM fields, we have obtained an expression for the energy stored in the host medium due to the scattering by inclusions.
This expression, from which we have calculated the bulk dwell-time and the quality factor of the system, has allowed us to achieve a parameter regime of enhanced field intensity in the host medium.
 Most importantly, the condition to achieve an enhanced quality factor is a packing fraction equal to the ratio between transport and scattering mean free paths, and hence a negative asymmetry parameter.
As a result, under certain conditions, disordered media with predominant backscattering response can exhibit a large quality factor, which could be of interest for lasing applications.  

\section*{Acknowledgments}

TJA holds grants from Coordena\c{c}\~ao de Aperfei\c{c}oamento de Pessoal de N\'ivel Superior (CAPES) (Grant No. 1564300) and S\~ao Paulo Research Foundation (FAPESP) (Grant No. 2015/21194-3), and
ASM holds grants from Conselho Nacional de Desenvolvimento Cient\'{\i}fico e Tecnol\'ogico (CNPq) (Grant No. 307948/2014-5).
FAP  acknowledges The Royal Society-Newton Advanced Fellowship (Grant No. NA150208), CAPES (Grant No. BEX 1497/14-6), and CNPq (Grant No. 303286/2013-0) for financial support.


\begin{thebibliography}{99}

\bibitem{ishimaru}
A. Ishimaru,
{\it Wave propagation and scattering in random media}
(Academic Press, 1978).

\bibitem{bohren}
C. F. Bohren and D. R. Huffman, 
{\it Absorption and Scattering of Light by Small Particles} 
(Wiley, 1983).

\bibitem{tiago-pra2016}
T. J. Arruda, A. S. Martinez, and F. A. Pinheiro,
``Electromagnetic energy and negative asymmetry parameters in coated magneto-optical cylinders: Applications to tunable light transport in disordered systems,''
Phys. Rev. A {\bf 94}, 033825 (2016).

\bibitem{bott}
A. Bott and W. Zdunkowski,
``Electromagnetic energy within dielectric spheres,''
J. Opt. Soc. Am. A {\bf 4}, 1361-1365 (1987).

\bibitem{tiago-sphere}
T. J. Arruda and A. S. Martinez,
``Electromagnetic energy within a magnetic sphere,''
J. Opt. Soc. Am. A {\bf 27}, 992-1001 (2010).

\bibitem{tiago-cylinder}
T. J. Arruda and A. S. Martinez,
``Electromagnetic energy within a magnetic infinite cylinder and scattering properties for oblique incidence,''
J. Opt. Soc. Am. A {\bf 27}, 1679-1687 (2010).

\bibitem{tiggelen}
A. Lagendijk, B. A. van Tiggelen,
``Resonant multiple scattering of light,''
Phys. Rep. {\bf 270}, 143-215 (1996).

\bibitem{pinheiro}
F. A. Pinheiro,
``Statistics of quality factors in three-dimensional disordered magneto-optical systems and its applications to random lasers,''
Phys. Rev. A {\bf 78}, 023812 (2008).

\bibitem{lacoste}
D. Lacoste and B. A. van Tiggelen
``Coherent backscattering of light in a magnetic field,''
Phys. Rev. E {\bf 61},  4556-4565 (2000).

\bibitem{albada}
{B. A. van Tiggelen, A. Lagendijk, M. P. van Albada, and A. Tip},
``{Speed of light in random media},''
{Phys. Rev. B} {\bf 45}, 12233-12243 (1992).

\bibitem{ruppin-energy}
R. Ruppin,
``Electric and magnetic energies within dispersive metamaterial spheres,''
J. Opt. {\bf 13}, 095101 (2011).

\bibitem{tiago-joa}
T. J. Arruda, F. A. Pinheiro, and A. S. Martinez,
``Electromagnetic energy within coated spheres containing dispersive metamaterials,''
J. Opt. {\bf 14}, 065101 (2012).

\bibitem{tiago-active}
T. J. Arruda, F. A. Pinheiro, and A. S. Martinez,
``Electromagnetic energy within single-resonance chiral metamaterial spheres,''
J. Opt. Soc. Am. A. {\bf 30}, 1205-1212 (2013).

\bibitem{tiago-coated-cylinder}
T. J. Arruda, A. S. Martinez, and F. A. Pinheiro,
``Electromagnetic energy within coated cylinders at oblique incidence and applications to graphene coatings,''
{J. Opt. Soc. Am. A} {\bf 31}, 1811-1819 (2013).

\bibitem{nieto}
M. Nieto-Vesperinas, R. G\'omez-Medina, and J. J. S\'aenz,
``Angle-suppressed scattering and optical forces on submicrometer dielectric particles,''
J. Opt. Soc. Am. A {\bf 28}, 54-60 (2011).

\bibitem{gomez}
R. G\'omez-Medina, L. S. Froufe-P\'erez, M. Y\'epez, F. Scheffold, M. Nieto-Vesperinas, and J. J. S\'aenz,
``Negative scattering asymmetry parameter for dipolar particles: Unusual reduction of the transport mean free path and radiation pressure,''
Phys. Rev. A {\bf 85}, 035802 (2012).

\bibitem{landau}
L. D. Landau and E. M. Lifshits,
{\it Electrodynamics of Continuous Media}
(Pergamon, 1984).

\bibitem{ruppin}
R. Ruppin,
``Electromagnetic energy inside an irradiated cylinder,''
J. Opt. Soc. Am. A {\bf 15}, 1891-1895 (1998).

\bibitem{Grenfell}
T. C. Grenfell,
``A theoretical model of the optical properties of sea ice in the visible and near infrared,''
J. Geophys. Res. {\bf 88}, 9723-9735 (1983).

\bibitem{liu} W. Liu, A. E. Miroshnichenko, R. F. Oulton, D. N. Neshev, O. Hess, and Y. S. Kivshar,
``Scattering of core-shell nanowires with the interference of electric and magnetic resonances,'' 
Opt. Lett. {\bf 38}, 2621-2624 (2013).

\end{thebibliography}
\end{document}